\documentclass{article}
\usepackage{graphicx}
\usepackage{natbib}
\usepackage[margin=1in]{geometry}
\usepackage[labelfont=bf]{caption}
\title{Using economic value signals from primate prefrontal cortex in neuro-engineering applications }
\author{Tevin Rouse$^{1}$, Shira Lupkin$^{2}$, Vincent McGinty$^{1}$ }
\date{}

\begin{document}

\maketitle
\begin{enumerate}
    \item Center for Molecular and Behavioral Neuroscience, Rutgers University–Newark, Newark, NJ, USA.
    \item Department of Neurobiology, University of Chicago, Chicago, IL, USA.
\end{enumerate}
\section*{Abstract}
Neural signals related to movement can be measured from intracranial recordings and used in brain-machine interface devices (BMI) to restore physical function in impaired patients. In this study, we explore the use of more abstract neural signals related to economic value in a BMI context. Using data collected from the orbitofrontal cortex in non-human primates, we develop deep learning-based neural decoders that can predict the monkey’s choice in a value-based decision-making task. Out-of-sample performance was improved by augmenting the training set with synthesized data, showing the feasibility of using limited training data. We further demonstrate that we can predict the monkey’s choice sooner using a neural forecasting module that is equipped with task-related information.  These findings support the feasibility of user preference-informed neuroengineering devices that leverage abstract cognitive signals.

\section*{Introduction}
Decoding neural signals is a primary focus of neuroengineering applications. By extracting the neural correlates of a particular cognitive process, neuroengineering systems like brain machine interfaces (BMIs) can help restore or augment that process. For example, BMI systems can partially restore motor functions by decoding movement-related signals from cortical motor areas, allowing users to control robotic limbs (\cite{velliste2008cortical},\cite{oby2020intracortical},\cite{shenoy2003neural}) or on-screen cursors (\cite{irwin2017neural},\cite{kim2008neural},\cite{gilja2012high},\cite{jarosiewicz2015virtual}),\cite{sussillo2012recurrent},\cite{fan2014intention}). Recent work has expanded on these ideas to decode neural correlates of sensory signals (\cite{gao2014visual},\cite{guenther2009wireless},\cite{hill2004auditory},\cite{jia2017decoding},\cite{o2011active},\cite{heelan2019decoding}). Other work in non-human primates has focused on invasive BMI applications for the prefrontal cortex to decode signals related to eye movements and spatial locations in a virtual maze (\cite{johnston2023decoding}). Despite these advances, very few studies have explored the use of abstract cognitive signals related to subjective mental contents (\cite{card2024accurate},\cite{wairagkar2023synthesizing},\cite{chinchani2022tracking},\cite{even2017augmenting}) . By detecting signals that are related to a user’s goals or beliefs, systems can be developed to aid higher cognitive functions.

In this work, we seek to explore the development of BMI systems that can extract abstract neural signals encoding economic values. Economic value (also called utility) is the subjective perception of the benefit or harm associated with a given stimulus, outcome, or course of action. Values are typically learned through experience and depend on an individual’s behavioral context, goals, and internal state.  Neural representations of value form the foundation for goal-directed decision-making (\cite{castegnetti2021usefulness}, \cite{o2011contributions}). Thus, the ability to extract subjective value information from neural activity is an important step in decoding abstract mental contents such as goals and desires. The neural correlates of value have been extensively characterized in animal models including nonhuman primates (\cite{mcginty2023behavioral},\cite{sharma2024orbitofrontal},\cite{padoa2006neurons},\cite{padoa2017orbitofrontal},\cite{thorpe1983orbitofrontal},\cite{conen2015neuronal},\cite{kimmel2020value},\cite{padoa2013neuronal},\cite{mcginty2016orbitofrontal},\cite{kennerley2009neurons},\cite{yamada2018free},\cite{rich2016decoding},\cite{padoa2006neurons},\cite{tremblay1999relative}).  Thus, this work seeks to operationalize neural representations of subjective value to explore the development of value-based BMI systems.

Much of the work on the neural representation of subjective value has focused on the orbitofrontal cortex (OFC), a prefrontal cortical region that has been shown to be critical for value-driven behaviors. OFC neurons not only encode economic value, but the disruption of this encoding disrupts many kinds of decision-making behavior, in both humans and animal models (\cite{rudebeck2008amygdala},\cite{rudebeck2011dissociable},\cite{torregrossa2008impulsivity}). Given the role of OFC in the decision-making process, our objective is to assess whether OFC-derived signals can be used in a BMI context to help the user achieve their goal. In this study we use offline computational approaches to decode subjective values and intended choices from the activity of OFC neural ensembles, using an existing data set from monkeys performing a simple economic decision task. We use the performance of our neural choice decoder to gauge the potential of a BMI that relies on OFC-derived value signals. Because value signals are context-dependent and dynamic, we use a reinforcement learning approach in developing a value-based neural decoder.  This framework permits updating the decoder with changes in a participant's subjective evaluations.

We will explore two challenges that need to be addressed to realize a value-focused BMI system using a deep learning framework: identifying neural network architectures that can potentially anticipate a user’s goals, and limited trial counts available for training the system. We will begin our assessment of BMI feasibility from the OFC by outlining a value-based two-alternative forced choice scenario. We will then demonstrate, using a transformer encoder-based agent (i.e. a neural decoder), the ability to extract value-related information from neural data recorded in nonhuman primates, and to use this information to predict choice behavior. Using this framework, we also explore the capabilities of a system that can make faster predictions by forecasting neural activity using multi-modal information within a decision-making context. We observed that we could obtain above-chance predictions on which item the monkey desired, including in cases where neither item was objectively better. We also observed that using forecasted neural activity allowed the decoder to make choice predictions sooner, but only when other information about the task conditions (i.e., the stimuli present in each trial) was incorporated into the forecasting process.

\section*{Methods}
\subsection*{Problem Description}
\begin{figure*}[h!]
\centering
\includegraphics[width=.9\linewidth]{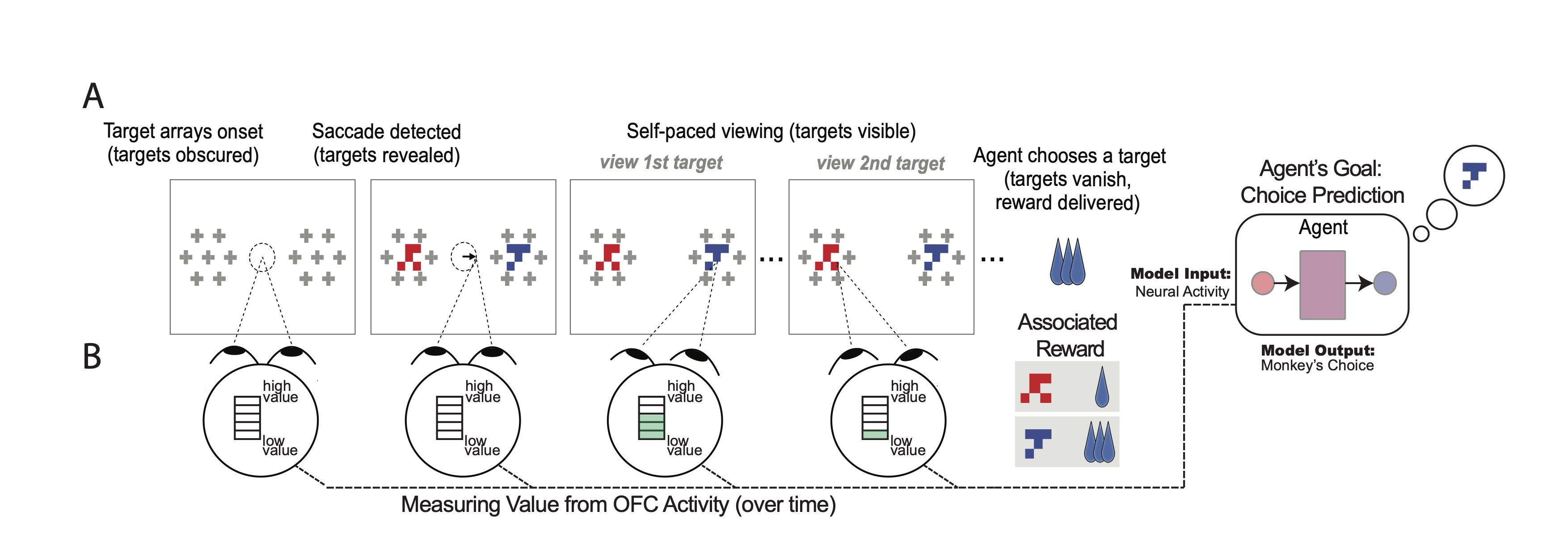}
\caption{ \textbf{Economic value-based BMI (brain machine interface) framework.} \textbf{(A)} Shown is a two-alternative forced-choice economic decision task that the monkeys performed while OFC neural activity data was collected. For task and behavior details see \cite{lupkin2023monkeys}. The monkey initiates a trial by fixating on a dot in the center of the screen, after which two covered stimuli appear. When the monkey saccades towards one of the stimuli, the stimulus in the direction of the saccade is revealed. Crowders (i.e., the visual elements surrounding each stimulus) force the monkey to look at the stimulus to gain information from it; they cannot identify the stimuli using peripheral vision. Each stimulus is associated with a reward ranging from 1-5 drops of juice; physically distinct stimuli can have the same reward association, meaning that in some trials there is no objectively better (largest reward) choice. Monkeys press a lever to indicate their choice, after which they receive the associated juice reward. \textbf{(B)} We consider a hypothetical scenario where a neural decoder extracts information from the neural activity obtained from the monkey while it is examining the stimuli. The decoder is tasked with extracting value information related to the stimuli and predicting the monkey’s choice. See Fig. 2A and Fig. 4A for model architectures tested in this study.}
    \label{fig: Fig1}
\end{figure*}
The objective of this study is to design a value-based BMI system that predicts economic decisions. We tested this design by building a neural decoder that uses the neural activity collected from monkeys performing a two-alternative economic decision task (a subset of the data used in \cite{mcginty2023behavioral}). On each trial, the decoder takes a multivariate time series (i.e., spike count data from a group of cells over time) and extracts the relevant value signals needed to predict the monkey’s choice. To explore the effect of the decoder design on prediction performance, we consider two different decoder architectures, detailed below.  All modeling was done using Pytorch 2.1.1 and scikit-learn 1.2.0.

\subsection*{Behavioral Task}
We use data from \cite{mcginty2023behavioral}, which used the following two-alternative forced-choice task to understand the neural mechanisms of value-based decision making. In this task, monkeys were tasked with examining two items on a display and selecting one (Fig. 1A). Each item, upon it being chosen, yields a fixed amount of juice reward of 1, 2, 3, 4, or 5 drops. On each trial, the monkeys were allowed to examine each option as long as they wanted, and to choose when ready. When the reward associations of the two items differed by 2 drops or more, the monkeys almost always chose the larger item; but when they were equal or near-equal, the monkeys’ decisions were variable from trial-to trial (see \cite{lupkin2023monkeys}, their Fig. 1D).
\subsection*{Data and Preprocessing}
We use 23 sessions of neural data collected from two rhesus macaques performing the value-based decision-making task in Fig. 1A (Monkey C: 10 sessions, Monkey K: 13 sessions); this is a subset of the data reported in \cite{mcginty2023behavioral}.For this work we only used cells that were identified to be in the posterior portion of the orbitofrontal cortex. We excluded sessions with less than 10 cells and fewer than 400 trials as well as sessions that contained $Na$ values. All analyses were performed within each session’s data. For a given session, the neural data consists of spike count data collected from single neurons in the orbitofrontal cortex (OFC, mean of $27.6$ SEM $2.7$ cells per session). Spikes from single cells were counted in $50$ milliseconds (ms), non-overlapping time bins. Given the structure of the task, we examined data time locked to when the monkey viewed the second target in each trial, which is the time point at which the monkey has acquired all the information necessary to make a decision (i.e. the identities of both targets). We used spike count data from $200$ ms before viewing the second target to $400$ ms after viewing the second target. The neural data in each session had the following structure: Trials x (Number of Cells) x (12 Time Bins of 50ms each). The behavioral data includes the choice that the monkey made for each trial along with the identity of both stimuli.

Generalizability for the trained agent was assessed by a series of random train-test splits, in which the fraction of training trials was varied between 20$\%$ to 80$\%$ in increments of 10$\%$. For each session we perform only one train-test split at each of the training fractions tested. We performed this data split randomly since we had no reason to maintain the original trial order of the session. Since data was collected after the monkey was trained on a stimulus set (see \cite{lupkin2023monkeys}, for details), we would not expect within-session learning effects if we had maintained the original trial order. Before model fitting and testing, we preprocessed the neural data as follows: for each cell and time bin in the series, we mean centered and scaled the data to unit variance, and then scaled the data again to the interval $[-1,1]$. The statistics used in the scaling process were derived for each cell using the portion of the data used in training.

\subsection*{Reinforcement Learning Approach}
An agent (i.e. the neural decoder) was trained with an environment constructed from collected neural data. The agent was tasked with taking neural data as input and predicting which of the two stimuli the monkey chose on a given trial. (Note that choices were variable when the two targets indicated equal or near-equal rewards, see \cite{lupkin2023monkeys} for details.)

While there are several supervised approaches that could be used to train the agent, in this study we use a reinforcement learning (RL) approach because an RL-based agent could more easily be extended to accommodate the dynamic, context-dependent nature of many value-based choices in natural settings. For further details, see Discussion. In the RL setting used here, an environment consists of a series of states. Because our environment is built from the trials performed by the monkey, each state corresponds to a trial. At each state the agent would be provided spike count data from a single trial.

The agent must then use the neural activity from the trial to predict whether the monkey chose the first item viewed or the second item viewed. When the agent’s choice matched the monkey’s choice (i.e. a correct prediction), it received a feedback signal from the environment of $1$, and received $0$ otherwise. The agent was not provided with information about which item was viewed first, or which part of the neural activity corresponded to the first- or second-viewed item; thus, the agent must learn to extract the relevant information from the time course of each trial.

Given that each session’s data was limited (mean of 631.4 trials), we added uniform noise (bound of $0.1$ around $0$) on each trial to create synthetic trial data to increase the size of the training set (\cite{saiz2005ensemble},\cite{holmstrom1992using}). This noise was added to the scaled data (see Data Preprocessing section) and clipped to maintain the $[-1,1]$ value range. We explored values for the noise’s bounds between $0.1$ and $0.2$ but these variations were quantitatively similar.

Training was organized into episodes. Each episode consisted of a set of training trials in which the agent uses the neural signal to render a prediction about the monkeys’ choice and receives feedback about whether the prediction was correct. Thus, each training trial consists of a time series of activity from the recorded cells, a predicted choice, and a feedback signal. The neural time series used for training were synthesized from the original data as described above, with the number of unique synthesized training trials in each episode matching the fraction of trials used for training, which varied between $20\%$ and $80\%$. Each subsequent episode used a different set of synthetic training trials. The agent completed $250$ of these episodes, at which point we sampled a subset of $10,000$ trials across all episodes and used these $10,000$ trials to update the agent’s parameters. The process above was repeated for $300$ iterations. Once the agent’s performance converged, we examined the agent’s choice prediction capabilities on held-out data absent added noise. All analyses related to the agent’s performance are evaluated in the noiseless case. All agent models were trained using a $Q$ learning approach which is estimated by the equation $Q(s,a) = r + g * max_{a^{'}} (s^{'},a^{'})$. In this equation g corresponds to the discount factor. Because in this task there are no inter-trial dependencies (the decision on trial t does not bear on trial $t+1$), we only considered a discount factor value of $g=0$. We used the mean squared error (MSE) as the loss function and Adam optimizer with an initial learning rate of $1e-4$ for parameter updating.

For agents that use an Internal Model to forecast neural activity (see Model Architecture below, results in Figs. 4 and 5), we used a two-step training process. The Internal Model is a type of variational autoencoder and so can be trained to reconstruct the input data. In the first step, the Internal Model (a Stochastic Recurrent Neural Network, SRNN) was trained to reconstruct neural activity using only the training data (i.e. not the held-out data). At each time step $t$, the ELBO (evidence lower bound) loss was computed. We trained this autoencoder minimizing the negative sum of the ELBO loss terms computed at each time t.

After we trained the Internal Model, the decision-making agent was trained as described above with the internal SRNN model weights held fixed (i.e. frozen). With the Internal Model held fixed, the agent uses this model to reconstruct the provided neural data and rely on permitted portions of the reconstructions to make choice predictions (see Model Architecture section for details on data masking). The training process was performed until the agent’s performance converged. After training the internal-model-equipped agents, we consider two ways to examine its performance. The first approach (Fig. 4) requires the agent to make choice predictions using the reconstructions of the provided neural data produced by the Internal Model. This performance evaluation is equivalent to the conditions used in the training process. Performance values derived from this analysis will be considered as a performance upper bound for the agent. The second approach (Fig. 5) requires the agent to first forecast neural activity and then use this forecasted data to make choice predictions. By forecasting we mean that neural data predictions made by the Internal Model on an upcoming part of the trial are influenced by its neural data predictions on previous parts of the trial. Performance under this approach is closest to that of a hypothetical device providing assistive signals to a user in real time. To assess how much the Internal Model contributes to the agents performance, we trained a version of the Core model (see Model Architecture section for details) in the same manner as the internal-model-equipped agent. We will refer to the Core model trained in this manner as the BASE agent/model (Fig. 5).

\subsection*{Model Architecture}
We considered two model variants. The objective of both models was to process multivariate time series data and extract the relevant neural signals needed to predict the monkey’s choice.

\underline{\textit{Core model variant}} The first model variant used a single transformer encoder module to interpret the neural data. We refer to this model as the “Core model/agent”. The architecture for the “Core model” (Fig. 2) consisted of first embedding the data using a linear layer and then applying positional encoding (using a sine function) to maintain the temporal ordering. Using a Bert style approach (\cite{devlin2018bert}), the output along with a class token is then fed into a 2-layer transformer encoder. The first dimension (i.e., the dimension corresponding to the class token position) is then extracted and used for computing the Q values corresponding to the possible actions that the agent could take. The layer that outputs the predicted Q values contains two linear layers with an instance normalization layer between them.  The action with the largest Q value is the one taken by the agent.

\underline{\textit{Internal Model variant}} The second variant is like the Core model but also contains an environment model (referred to as “Internal Model” in Fig. 4) whose output is interpreted by a second transformer encoder. The Internal Model provides the agent with a forecast of future neural activity based on the neural activity observed up to a time point $t < t_{max}$ and information about the trial condition. Given the structure of the task, we deviated from a modern characterization of an environment model which requires future state predictions dependent on the agent’s choices (\cite{moerland2023model}). Because trials consist of only a single choice and states correspond to neural activity, we considered an environment model that predicts the neural activity that will appear in upcoming $50$ millisecond time bins within a trial. We used the SRNN (\cite{fraccaro2016sequential}) architecture for this purpose (see above). Given below are the underlying expressions that represent the forecasting module. We use the notation provided in (\cite{girin2020dynamical}) to describe the original model (\cite{fraccaro2016sequential}). Let $x_{t-1}$ be the neural activity and $h_{t-1}$ be the output from the hidden layer at time point $t-1$. To compute the output from the hidden layer at time $t$, we provide as input both $x_{t-1}$ and $h_{t-1}$ into the recurrent layer represented by $d_{h}$. Deviating from the original model, we provided a stimulus vector $s_{t}$, representing the identity of the target available at time $t$. We concatenated this stimulus vector to the input vector $x_{t}$ and provided this composite representation as the input to the recurrent layer $d_{h}$.
\begin{equation}
    h_{t} = d_{h}(concat(x_{t-1},s_{t-1}),h_{t-1})
\end{equation}
Because the forecasting module is a variational autoencoder, we leverage the output from the hidden layer to approximate the mean $\mu$ and standard deviation $\sigma$ of the distribution from which we sample the latent vector $z$. We assume that the underlying conditional probability distribution P estimated using a model parametrized by $\theta$ is a multi-variate normal distribution.
\begin{equation}
    P_{\theta_{t}}(z_{t}|z_{t-1},h_{t}) = N(z_{t};\mu_{\theta_{z}}(z_{t-1},h_{t}),diag\{\sigma^{2}_{\theta_{z}}(z_{t-1},h_{t}) \})
\end{equation}
To compute the mean and standard deviation for this distribution at time $t$, we provide the latent vector (derived using the reparameterization trick) from time $t-1$ and the output from the hidden layer at time $t$. Both the latent vector at time $t-1$ and the hidden layer’s output at time $t$ are provided to the function $d_{z}$ which represents the encoder component of the autoencoder.
\begin{equation}
    [\mu_{\theta_{z}}(z_{t-1},h_{t}),\sigma_{\theta_{z}}(z_{t-1},h_{t})] = d_{z}(z_{t-1},h_{t})
\end{equation}
To compute the predicted neural activity at time $t$, we provide the decoder $d_{x}$ the latent vector $z_{t}$ and $h_{t}$. The decoder produces the predicted neural activity $x_{t}$.

The amount of actual neural data provided to the model (and therefore the proportion of actual vs. forecasted data used to make predictions) was varied across trials.  To train the agent over these variable temporal components, we leveraged the masking mechanism present within the transformer encoder architecture to vary what time points the agent had access to on a particular trial. For each state, the agent is provided with a multi-variate time series and two masking vectors. These vectors determine which time bins the agent can attend to make its choice prediction. To encourage the agent to learn how to extract information from both modules, we designed the mask so that the forecasting module would only attend to forecasted time points while the core module attended to a part of the spiking data that was not forecasted. One vector is given to the transformer encoder present in the core module. The other vector is given to the transformer encoder present in the forecasting module. For example, consider a trial in which only the first 6 of the 12 total time bins of neural data are available. In such a trial, the mask for the core module permits the agent to attend to the first 6 time bins, while the agent’s forecasting module would predict the next 6 time bins and attend to only those 6 predicted bins. Once both modules have evaluated their inputs, the representations from each path are combined and used to make a choice prediction.

We randomly varied the parts of the time series that are masked but in all cases the masks were continuous blocks of time, consistent with the goal of designing an agent to predict choices based on the activity observed up to a given time point. We never consider providing masks to the agent in which random non-consecutive time bins were masked. Given the variability in the environment, we report performance metrics for each neural recording session that was averaged over 20 simulations. We report the distribution of the averaged performance metrics. Along with the states and masking vectors we also considered two cases: (1) An agent provided with information about the targets and (2) an agent that is not provided with information about the targets. In case (2) the agent is given the neural data and masks as described above and is required to reproduce (or forecast in post-training scenarios only; Figure 5) neural activity. In case (1) the agent can incorporate information about the targets viewed in its prediction of the upcoming neural dynamics.

To provide the agent with target information, we provided the Internal Model with a unique 50-dimensional fixed tag associated to each stimulus provided during the trial. This allows the Internal Model to condition its forecast of orbitofrontal cortex activity on the presented stimuli. We provided the Internal Model with the 1st stimulus vector when the portion of real neural activity occurred before the 2nd fixation. We then provided the 2nd stimulus vector for time windows that occurred after the 2nd fixation.Combining the neural activity forecast from the Internal Model with the core module architecture allowed us to consider choice predictions at earlier points in the trial. After the initial training of the Internal Model, the Internal Model’s weights remained fixed while the agent’s weights were allowed to update as described above.

In summary, the model (Fig. 4A) contains two processing modules that aid in choice prediction: (1) a module that uses neural activity observed up to a given time point and (2) a module that forecast future dynamics of the trial and uses only the forecasted activity. Both the current trial information and forecasted trial information combined and used to predict the Q value. See Figure 4A for a diagram of the architecture. Representations for both paths were done using a 2-layer transformer encoder module with the internal feature dimensionality of 100 and RELU (rectified linear unit) activation for the intermediate layers.

\subsection*{Agent Analysis}
To quantify the amount of value information present in the internal representations of the agent, we decoded the associated value of the stimuli present in the trial using the agent’s internal representations. We performed this analysis using the Core model (see Model Architecture section). For each trial we extracted the 100-dimensional vector corresponding to the first dimension of the transformer encoder module output. This vector was aligned to the position of the class token. With this collection of vectors, we used linear regression to decode the value of the stimuli. Decoding was conducted using the $80\%$ training data case introduced in Fig. 2. The linear regression was fit to the representations generated from the training data and tested on the representations derived from the held-out data. The violin plots shown in Fig. 3A show the distribution of $R^2$ values over all neural recording sessions.

Given the difference in data preprocessing used in this work compared to the previous study (\cite{mcginty2023behavioral}) we verified the ability to decode both targets’ value (results in Fig. 3B).  Value decoding was done with a linear regression model using the original neural data at each time bin. The decoding results reported in Fig. 3B were derived from 5-Fold CV with 10 repeats. To assess what time bins within a trial are important for the agent to predict the choice of the monkey, we performed Shapley analysis on each model(\cite{scott2017unified}). This analysis was only conducted on the held-out data. This was done using the Gradient Explainer function from the SHAP toolbox (\cite{scott2017unified}) in Python. To assess the magnitude of each time bin’s contribution within a recording session, we computed the average of the absolute value of the estimated Shapley values.

\section*{Experiments and Results}
\subsection*{Reinforcement learning approach for choice decoding }
Our goal is to develop an adaptive decoder that can predict the choices of monkeys in single trials of a two-alternative economic decision task. We trained this decoder using a collection of sessions, each consisting of hundreds of trials of choice behavior data and concurrently observed activity in OFC neurons. Because the data comes from data collected on different days, each session has neural activity from a distinct set of cells varying from session to session. Because cells in different sessions have different firing patterns, it is not feasible to pre-train a model on data from one session and test it on data from another. To circumvent this issue, we split the trials randomly within a session into a set used for training the model and a set used to measure how well the decoder's performance extends to new trials. The fraction of trails used for training varied from 20 to $80\%$.

To train the neural decoder, we used a reinforcement learning approach (\cite{pohlmeyer2014using},\cite{sanchez2011control},\cite{digiovanna2007brain},\cite{digiovanna2008coadaptive},\cite{bae2015kernel}). Within an RL viewpoint, a neural decoder can be viewed as an artificial agent interacting in an environment defined by neural activity. The agent’s goal is to reproduce the choice that the monkey made in each decision trial given the neural activity. To guide how model parameters are updated, our environment yields a feedback signal (i.e., reward) of 1 when the agent’s choice matches the choice made by the monkey and 0 when it does not. Our definition of the feedback signal deviates from how RL approaches are used in neural decoding applications because it is not derived from an internal signal (\cite{marsh2015toward},\cite{prins2014confidence}).

\begin{figure*}[h!]
\centering
\includegraphics[width=.9\linewidth]{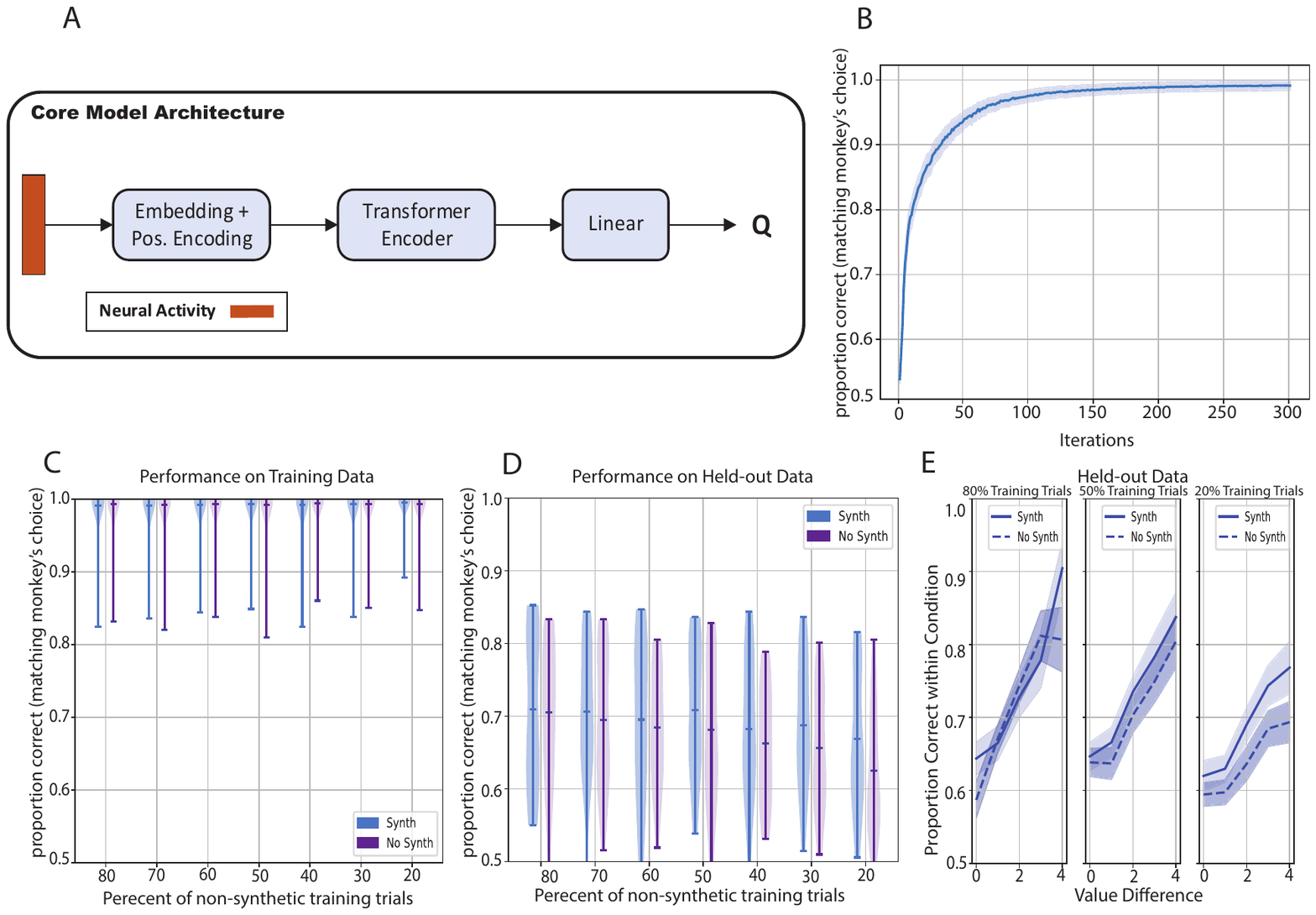}
\caption{\textbf{The reinforcement learning agent can learn to predict the monkey’s choice.} \textbf{(A)} Shown is the Core model architecture used for the agent. The agent model is designed to take as input neural activity (spike counts over time) and predict the Q-values associated to each state-action pair. \textbf{(B)} The proportion of trials in which the agent correctly predicts the monkey's choices over the training process. Shaded region corresponds to the session-wise SEM over n=23 sessions. \textbf{(C)} Distribution of the agent’s choice prediction performance on the training data at the end of training. Performance is reported on the case when the agent is trained with and without synthetic data (see Methods). The central bar in each violin plot corresponds to the mean performance over sessions while the bars on the edges correspond to the range. The x-axis gives the percentage of trials used in the training data for each session. There were no significant differences between Synth and NoSynth training approaches (Table 1). \textbf{(D)} Distribution of the agent’s performance on the held-out data over sessions. Plotting conventions are the same as in (C). Performance was higher when synthetic data was used in runs using $50\%$, $30\%$ and $20\%$ training data (Table 2). \textbf{(E)} Performance of the agent on the held-out data as a function of the value difference condition for the $80\%$, $50\%$, and $20\%$ training data cases with and without synthetic training data. Shaded region corresponds to session-wise SEM.}
    \label{fig: Fig2}
\end{figure*}

We trained the agent using a standard deep Q network (DQN) approach (\cite{barto2021reinforcement}).  Shown in Figure 2A is the architecture of the agent. Because the agent receives a multivariate time series as a state input, we equipped our agent with a network architecture that could process temporal dependencies. Given the performance capabilities demonstrated in studies tackling sequential data classification (\cite{dosovitskiy2021an}, \cite{burchi2023audio}), we chose to use a 2-layer transformer encoder with positional encoding to process temporal dependencies within the neural dynamics. We first examined the agent’s performance on the training data (Fig. 2B). After a few hundred iterations, the agent learned to predict the monkey’s choices in the training set with near-perfect accuracy.

Although this demonstrates that the neural decoder can learn to match the choice preferences of the monkey, we sought to answer two important concerns related to feasibility. The first concern relates to the ability of the decoder to generalize to new trials, which is a weakness in some RL models (\cite{ghosh2021generalization}). The second concern relates to the number of trials needed to train the model to make accurate predictions, given that in a BMI setting practical considerations limit the available training time. We jointly tested these concerns by testing decoder performance in held-out trials, and by varying the number of trials in the training and held-out splits (Fig. 2C, 2D). Given the relatively low trial counts (mean 623 trials per session), we augmented the training data by creating “synthetic” trials. The cell firing rates in synthetic trials are based on the original training trials but have uniform noise added to each of the cell firing rates (see Methods sections on data preprocessing and reinforcement learning approach). The noise was independent for each unit. The performances on training and held-out trials, shown in Fig. 2C and 2D, allow us to gauge how many non-synthetic trials are needed to train the decoder.

Performance on held-out data was lower than in the training set, but still above chance (Fig. 2D). When synthetic data were used for training, predictive accuracy degrades slightly (within 10 percent difference) as the percentage of training trials decreases from 80 to $20\%$ (Fig. 2D, light blue). This suggests that with as few as 20 percent of trials (average 126 per session) we can still obtain above-chance out of sample performance. In contrast, when synthetic data are not used in model training (Fig. 2D,dark blue), performance in held out data decreases significantly for the $20\%$ training set case (Fig. 2D, Synth vs. NoSynth, $p= 3.83e-4$ by t-test). Thus, the use of synthetic training trials decreases the number of training trials needed to obtain good decoder performance.

We next asked how performance depended on the difference in value of the two stimuli presented, which is an index of trial difficulty. Shown in Fig. 2E is the proportion of trials that the model correctly predicted for each value difference, using $80\%$, $50\%$, and $20\%$ of the nonsynthetic data trials as a training set (Fig. 2D, far left on the X-axis). We observed above-chance performance at all value differences. Above-chance performance on value difference 0 trials (when neither choice is objectively correct) is notable, because it implies that the agent can detect the monkey’s subjective sense of value as it fluctuates on individual trials. Decoding performance on the high value difference trials was far less than $100$ percent on average.  This contrasts with the monkeys’ choices in such trials, which were nearly $100\%$ in favor of the best option (e.g. \cite{lupkin2023monkeys}, their Figure. 1).  Thus, the information available to the decoder is not sufficient to completely replicate the monkeys’ behavior even for the easiest trials.  Consistent with Fig. 2D, in the $20\%$ training data case, the use of synthetic training data increased the prediction accuracy in held-out trials (Fig. 2E, right).

\subsection*{Interrogating the internal representations of the choice decoder}
\begin{figure*}[h!]
\centering
\includegraphics[width=.9\linewidth]{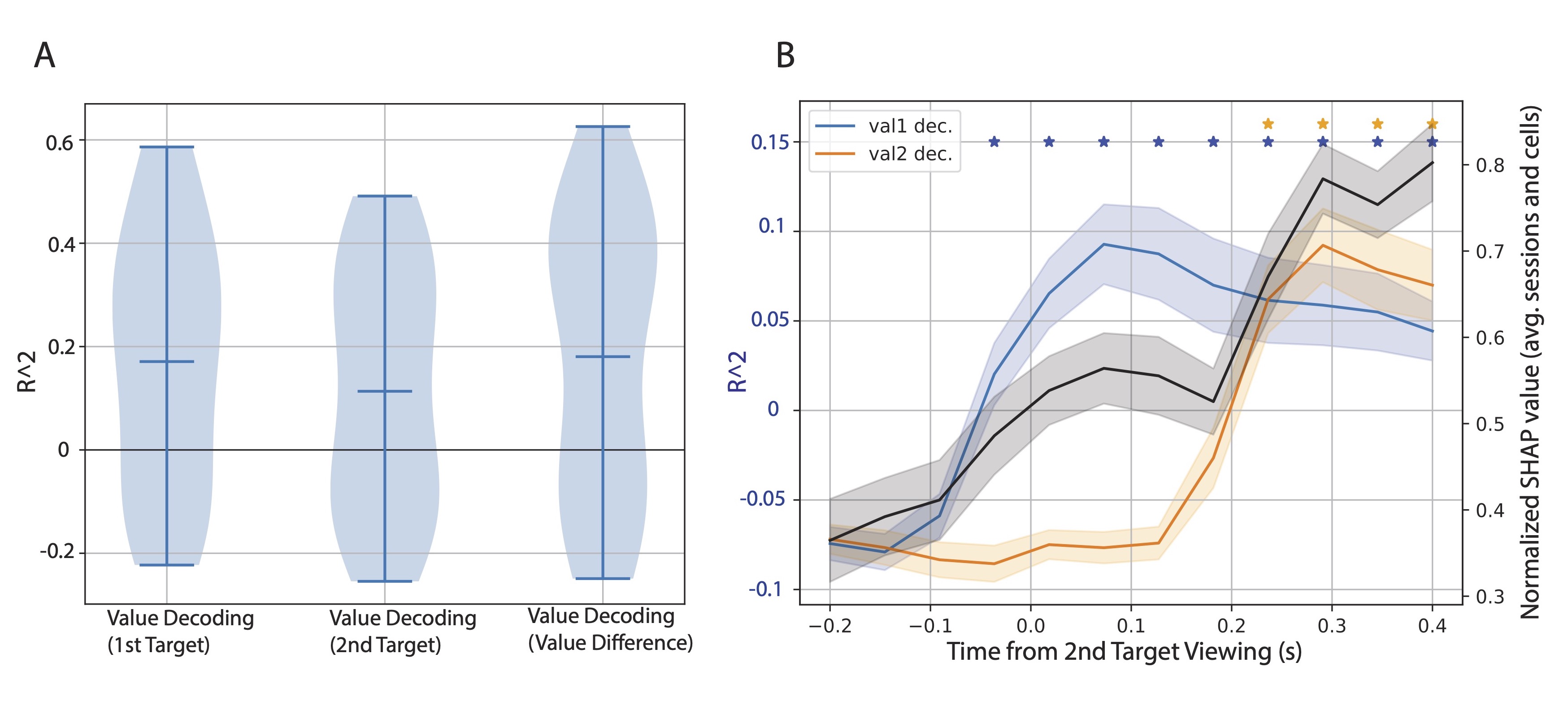}
\caption{\textbf{The agent predicts the monkeys’ choices using value information encoded near the times when each option’s value signal reaches a peak.} \textbf{(A)} Decoding performance ($R^2$) from the agent’s internal representation. A vector extracted from the transformer encoder’s output was used to decode the value of the 1st stimulus viewed, 2nd stimulus viewed, and the value difference between the 1st and 2nd stimulus. Decoding of each value was done using OLS applied to the representations generated from $80\%$ training data case. Shown is the distribution of $R^{2}$ values with the central bar representing the mean and the edges corresponding to the minimum and maximum $R^{2}$ value. Decoding performance of different value-related measures using the agent's internal representation (comparing the coefficient of determination to 0, $p< 0.05$ by t-test; Table 3). \textbf{(B)} The blue and orange curves give the decoding accuracy for the value of the 1st-viewed stimulus (blue) and second-viewed stimulus (orange) using the original neural data over time. The black curve shows the Shapley values for the agent’s $Q$ value predictions. The Shapley values were normalized within session and averaged over cells and trials. Shaded areas correspond to session-wise SEM. Stars indicate statistical significance for decoding value ($p < 0.001$).}
    \label{fig: Fig3}
\end{figure*}

Given the decoder’s prediction capabilities, we next asked what aspects of its input are important to making its decision. Previously we have shown that choice decoding from neural data could be performed at above-chance levels when using neural signals only related to the first of the two targets (\cite{mcginty2023behavioral}). Therefore, we asked whether the agent makes predictions using information about one target or both. We extracted the internal representations from the trained agent for each trial and attempted to decode the values of the two items in each trial. Because we use a transformer encoder, we use the output corresponding to the class token as the agent’s trial representation. We considered three variables in our decoding analysis: the value of the 1st target, the value of the 2nd target, and the difference between the 1st target value and 2nd target value. Shown in Fig. 3A is value decoding performance in held-out data, when using 80 percent of the data to train the model. We find above-chance decoding of value information from the agent’s representations for each target individually, as well as for the difference between the two targets’ values.

We next asked which time bins in the trial were important for the agent’s prediction, using a standard approach to Shapely analysis. Previous studies (\cite{mcginty2016orbitofrontal},\cite{hunt2018triple},\cite{mcginty2023behavioral}) have shown that value signals are dynamic and are computed time-locked to the viewing of the targets. These studies have also identified a delay of $~200$ ms between when the monkey sees the target and when value information is first detectable in OFC (E.g. \cite{mcginty2023behavioral}, figure 2).  Like these prior OFC results, we found that we could decode from the neural data the value of the first target in earlier time bins, and the second target value in later time bins (Fig. 3B, blue and orange lines). Considering these observations, we expect that later bins, i.e. after both targets have been viewed, will be more important for the agent’s $Q$ value predictions.

We examined the absolute value of the Shapley values to understand how much each feature (time bin) contributes to the Q-value estimates regardless of the sign. Over all sessions (Fig. 3B, black curve), the agent found later time bins ($>0.2$ after 2nd target viewing) to be the most informative. However, the results suggest that the agent also uses information from the early timepoints ($~0.1$s in Fig. 3B, when only information about the first target is available) instead of exclusively relying on the last time bins (when information is available for both targets). Together, these results indicate that the RL decoder predicts choices by extracting information about the values of the targets offered in each trial.

\subsection*{Augmenting Choice Decoding with Neural Forecasting}
\begin{figure*}
\centering
\includegraphics[width=.9\linewidth]{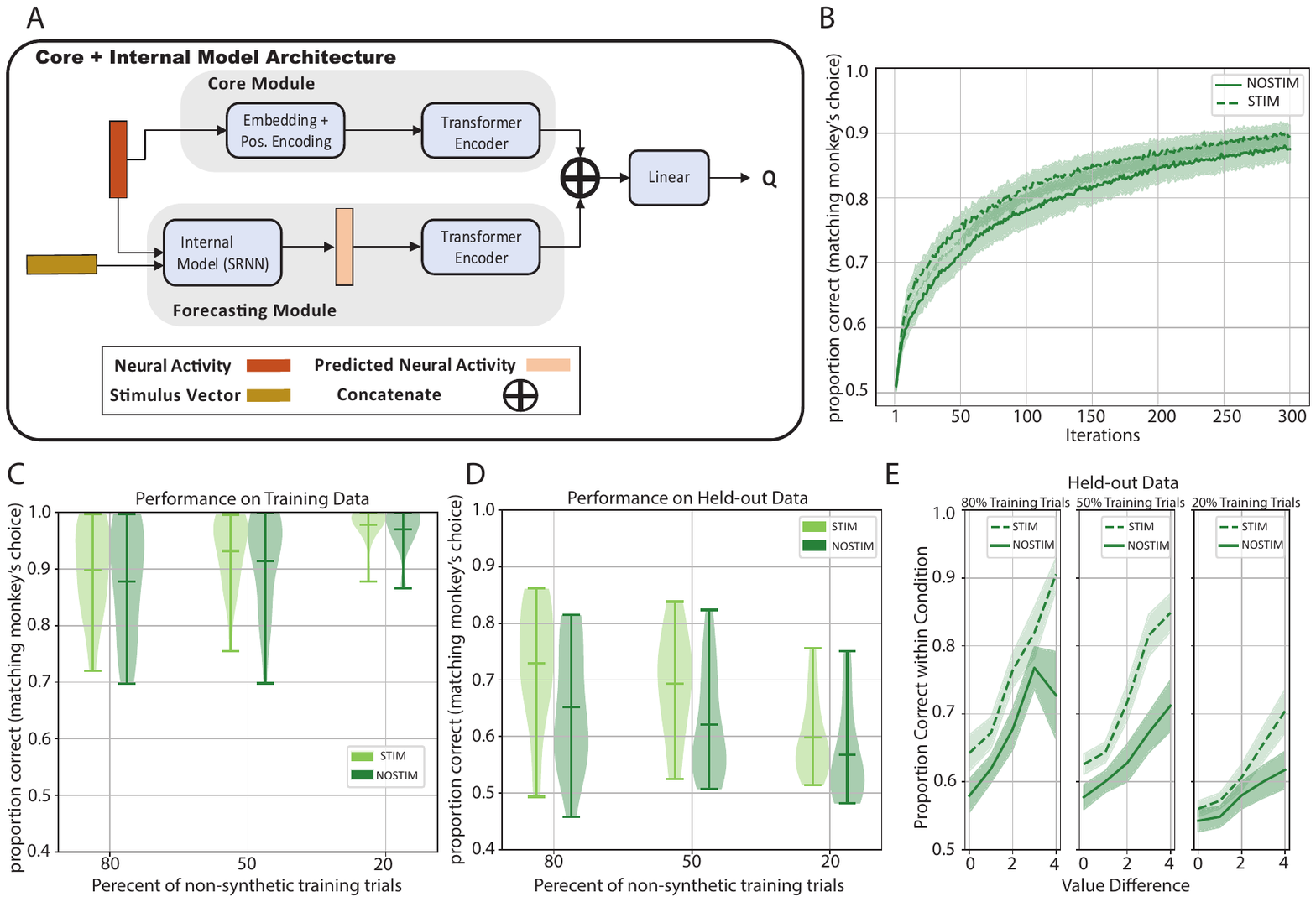}
\caption{\textbf{A reinforcement learning agent equipped with an internal forecasting model can learn to predict the monkey’s choice.} \textbf{(A)} The agent architecture consists of the same modules as the Core model (Fig. 2A) but also contains a processing structure for the internal forecasting model. The Internal Model receives both the OFC neural activity up to a given time point $t$ and vectors indicating the stimuli presented to the monkey; together, these are used to forecast the neural activity that would occur in OFC after time $t$. The agent then predicts the Q-value for both state-action pairs using the Internal Model's forecasts. For clarity, we will refer to the modules found in the Core model as the core module. The combination of the Internal Model and the transformer encoder that uses the Internal Model’s output will be referred to as the Forecasting module. \textbf{(B)} The performance (proportion of trials correctly predicted) of the agent over the training process. Note that during training and testing the amount of real vs. forecasted neural data (parameterized by the neural sampling time) $t < T$ was varied across trials; this is unlike the Core model (Fig. 2) which used the full sample of neural data in each trial ($t=T$). Performance is shown for agents trained without the stimulus vector (NOSTIM) and with the stimulus vector (STIM). Error-bars correspond to the session-wise SEM. \textbf{(C)} Distribution of the agent’s performance on the training data over sessions for the STIM and NOSTIM cases. The number of trials used in the nonsynthetic training data for each session was determined by a percentage of the total number of trials within that session. The central bar corresponds to the mean performance over sessions. The bars on the edges of the violin plots correspond to the minimum and maximum performance over sessions. Performance was significantly higher in the STIM condition (Table 4). \textbf{(D)} Distribution of the agent’s performance on the held-out data over sessions for the STIM and NO STIM cases. Plotting conventions are the same as those in (C). Performance was significantly higher in the STIM condition (Table 5). \textbf{(E)} Performance of the agent on the held-out data within value difference condition for the $80\%$, $50\%$, and $20\%$ training data case. Performance was quantified here as the proportion of trials correctly predicted within each value difference condition. Error-bars correspond to session-wise SEM. }
    \label{fig: Fig4}
\end{figure*}

Next, we extend the model above to consider an assistive BMI system that can anticipate a user’s abstract goals or intentions, and therefore guide the user to act more rapidly than would otherwise be possible. To achieve this objective, we take inspiration from human-robot interaction systems that aim to forecast the user’s actions before the robot makes a move (\cite{wang2013probabilistic}). Specifically, we explore a BCI system that is equipped with a neural forecasting module. In our hypothetical scenario, the agent is provided with a module that uses a portion of neural activity up to a given time point to forecast future neural activity and to then predict the monkey’s choice. Under the assumption that ongoing activity dynamics shape subsequent activity (\cite{remington2018flexible}), the forecasting module would permit the system to predict the OFC’s activity before it occurs, making faster goal inferences than would be possible from native neural data alone. In the robotics scenario, this would be akin to inferring the user’s intent slightly before the user’s motor execution, giving the system more time to plan how its actions will aid the user. Thus, we constructed an agent system like the one described above, but with an Internal Model whose function is to forecast future neural activity out to a maximum time T given the activity up to some point $t < T$ (Fig. 4A). The maximum time T in this work is 400 ms after the monkey viewed the second target (i.e. the full length of the time series).

Using this forecasting module, we also consider an additional method for increasing the agent’s speed and decision accuracy, by providing the Internal Model with information pertaining to the targets in each trial. Providing external information about the targets allows us to consider an assistive neural engineering system that uses or complements one's existing sensory capabilities to maximize the information available to (and therefore the predictions made by) the BMI agent. As described earlier, OFC’s activity depends on the visual targets presented in each trial, and the time course of OFC activity within a trial depends on when each target is viewed. Therefore, to more accurately forecast OFC activity, we provide the Internal Model with information pertaining to the targets in each trial, in the form of a fixed “stimulus vector” unique to each stimulus (Fig. 4A, brown box). Providing the agent with the information about the targets does not trivialize the problem because a sufficient portion of trials pertain to pairs of stimuli that possess the same value. In these trials there is no “objectively” correct choice, so that choices reflect the subjective preferences of the animal (\cite{mcginty2023behavioral}, \cite{sharma2024orbitofrontal}, \cite{lupkin2023monkeys}). Therefore, an important question is whether the agent’s predictive accuracy for these subjective choices is improved when stimulus information is available, or whether performance is improved only for choices with an objectively correct (higher-value) option.

Below, we assess the performance of agents using forecasting modules. First, we quantified the maximum benefit that a forecasting module could provide to the agent (i.e. choice-prediction performance upper bound;results in Fig. 4). We then quantified typical performance under conditions that most resemble our expected use case (results in Fig. 5).  In both cases, we compare the performance of agents with vs. without the stimulus vector information provided to the forecasting module.

\subsubsection*{Assessing the performance upper bound}

Given neural data up to a time point ($t_{0},..., t_{n}$), the forecasting module predicts the activity iteratively: It first predicts activity at $t_{n+1}$, and then uses the predicted data at $t_{n+1}$ to predict the activity at $t_{n+2}$, and so on until the end of the trial. Thus, any forecasting error occurring at $t_{n+1}$ propagates to subsequent time points into the entire forecasted portion of the data ($t_{n+1},...,T$), which ultimately affects the agent’s choice decoding performance. To estimate the system’s performance upper bound, we wished to minimize the effects of these time-propagated forecasting errors.

To do so, we allowed the Internal Model to provide the agent with reconstructed neural data, rather than forecasted data. Reconstructed data is obtained by providing the forecasting module with the real neural data prior to each time point. (Because the forecasting module is an autoencoder, it is optimized to reproduce (or “reconstruct”) the data provided as an input. Thus, the neural data reconstructions can be thought of as a neural data quality upper bound.) Using reconstructed data avoids the propagation of forecasting errors across timepoints and therefore minimizes the effects of the Internal Model derived errors on the agent’s choice prediction. We consider the performance when the agent uses reconstructions to make choice predictions (Fig. 4) as the highest performance (i.e. upper bound) achievable by this model.

We assessed the performance upper bound when the Internal Model has access to both neural activity and a vector representing stimulus information (STIM agent), as well as an Internal Model provided only with neural activity (NOSTIM agent). Both used synthetic data trials during training (i.e. as in the “Synth” models in Fig. 2). Across all the recording sessions, both STIM and NOSTIM agents achieved comparable performances on the training data (Fig. 4C). We next evaluated the generalizability of each agent by examining their predictions on held-out data (Fig. 4D).  In both model cases, both agent types saw a considerable performance drop as the fraction of training trials decreased.  Across all data partitions, the STIM agent outperformed the NOSTIM agent (Fig. 4C,D), and predictive performances were higher when the value difference was larger (Fig. 4E).  Thus, when neural forecasts are augmented by information about the target stimuli, the agent can more accurately predict the monkeys’ choices.

\subsubsection*{Forecasting neural data improves predictive accuracy early in a decision trial}
\begin{figure*}
\centering
\includegraphics[width=.9\linewidth]{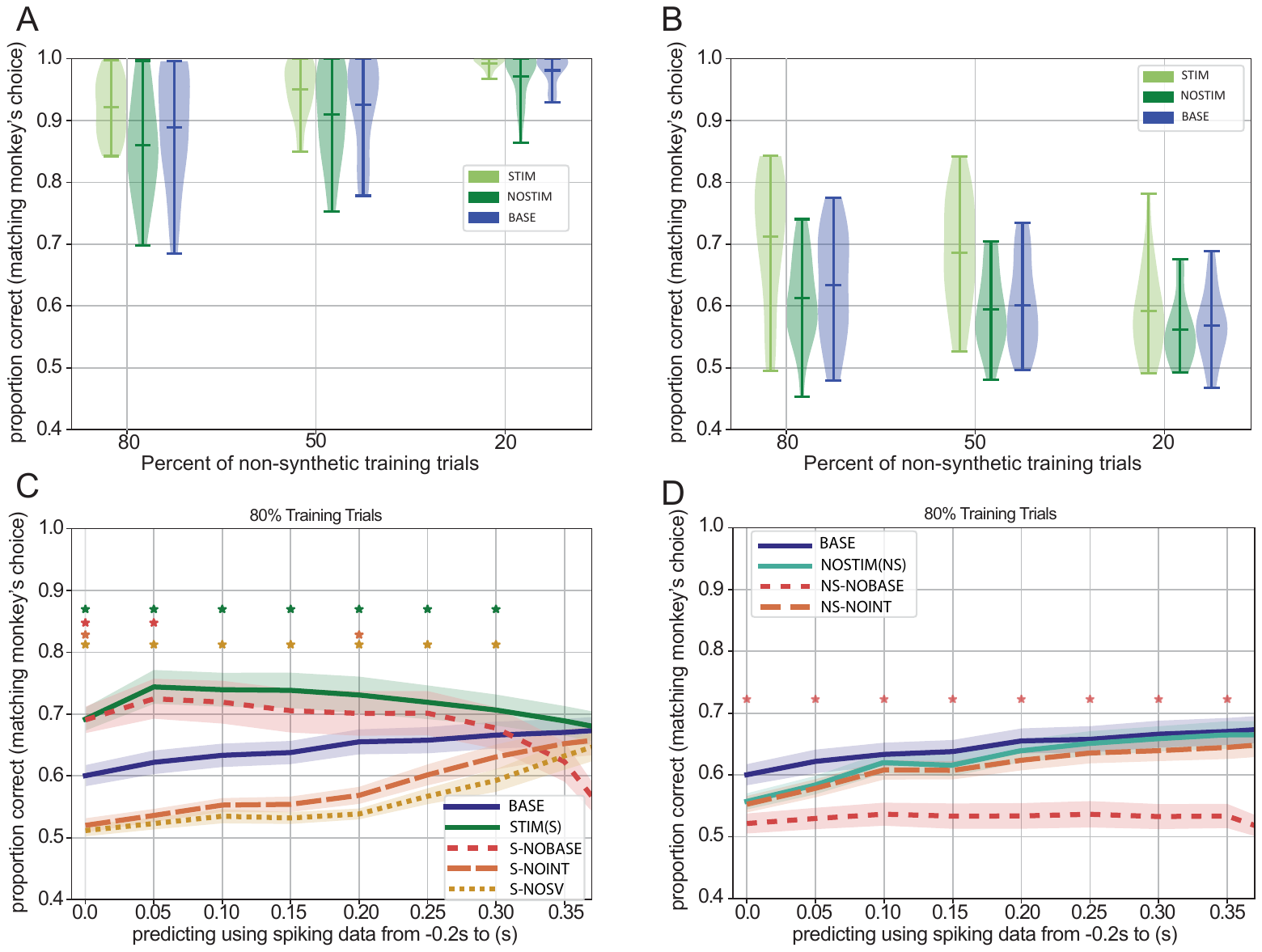}
\caption{\textbf{Internal Model The agent with forecasting module can predict choices when only a portion of each trial’s neural data is provided.} \textbf{(A)} Violin plots for each agent type over sessions. Performance on the training data is shown for the cases when the percentage of nonsynthetic data is $80\%$, $50\%$, and $20\%$ of the available trials within a dataset. All three agent types (i.e. STIM,NOSTIM,BASE) learn to make choices in an environment that varies the number of available time points the agent has access to. For statistical significance see Table 6 in Appendix. \textbf{(B)} Agent performances on the held-out data for the splits shown in (A). The STIM agent maintains superior performance on the held-out data compared to the other model variants. For statistical significance see Table 7 in Appendix. \textbf{(C)} Average performance (correct choice prediction) over sessions in the held-out data of the agent trained using only a portion of neural data ($80\%$ training data case).  The x-axis gives the upper limit (in seconds) of the neural data provided to the agent in each trial (the lower limit was $-0.2$s for all). The model cases within the legend: BASE: BASE agent (a version of the Core model which does not have a forecasting module but is trained in the environment that varies the number of available temporal components), STIM: Agent with forecasting module, with the stimulus vector provided during both training and testing, S-NOBASE: The STIM agent with the core module contribution removed during testing, S-NOINT:  The STIM agent with the forecasting module’s contribution removed during testing, S-NOSV: Forecasting agent when the STIM agent with the stimulus vector not provided during testing. Stars indicate statistical significance when comparing each case to the BASE model performance ($p < 0.001$). \textbf{(D)} Similar to (C). The model cases within the legend: BASE: Same as in (C), NOSTIM: Agent with forecasting module, with no stimulus vector provided during either training or testing, NS-NOBASE: The NOSTIM agent with the core module’s contribution removed during testing, NS-NOINT: The NOSTIM agent with the forecasting module’s contribution removed during testing. Stars indicate statistical significance when comparing each case to the BASE model performance ($p < 0.001$). }
    \label{fig: Fig5}
\end{figure*}

The results in Fig. 4 show the impact of the neural forecasting module under ideal conditions, specifically, when we eliminate time-propagated inaccuracies that originate from forecasting neural data (see Methods).  In contrast, Fig. 5 shows the impact of the forecasting module under more realistic conditions, specifically, when only an initial segment of neural data is provided, such that forecasting errors at each time step propagate into future time steps. To test this, we trained the agent using trials in which only an initial portion of real neural data is provided to the system (e.g. the first 6 out of 12 total time bins, variable across trials), requiring the Internal Model to forecast the remaining data. We trained agents both with and without stimulus information provided on every trial (STIM and NOSTIM, respectively). In addition, to assess the impact of the forecasting module, we trained an agent with no forecasting module (like the Core model in Fig. 2) to make choices in the same environment as the STIM and NOSTIM agents,(i.e. with a variable initial segment of neural data provided). We refer to this agent as the BASE agent.

In the training data, the STIM agent had the highest performance on average for all non-synthetic data splits (Fig. 5A). Despite not having access to a forecasting module, the BASE agent slightly outperformed the NOSTIM agent (see Table 6 for statistics), suggesting that forecasted activity lacking task-relevant information may negatively impact the training process. In the held-out data the STIM agent also outperformed the other agent types, with no significant difference in performance between the NOSTIM agent and the BASE agent (Fig. 5B, Table 7). These results suggest that the forecasting module provided no benefit to choice prediction in the absence of information about the task stimuli (i.e. the stimulus vector).

We next sought to characterize the impact of the forecasting module by assessing the agent’s performance as a function of how much initial neural data was provided (Fig. 5C). In addition, we evaluated the contribution of individual model components by selectively 'ablating' architectural components during testing (Fig. 5D). We focused our analysis on the $80\%$ training data case. The predictive performance of the BASE agent gradually increases as more data is provided (Fig. 5C, blue line), consistent with the observation that later time bins are more informative (Fig. 3B). In contrast, the STIM agent’s performance is high regardless of how much initial data is provided (Fig. 5C, green line), indicating how informative the Internal Model’s forecasted activity is for predicting choice even at early time points in the trial.

As we removed modules from the STIM agent, we observed complementary effects on performance. Removing the core module contribution had little effect in early portions of the time window but resulted in decreased performance at later time points in the trial (Fig. 5C, red line). A complementary pattern was observed after removing either the forecasting module or the stimulus vector: performance was near chance at early portions of the time window and improved at later time points (Fig. 5C, gold, and brown lines). These results demonstrate that the STIM agent depends mostly on the forecasting module at early time points and uses a combination of actual neural signals (from the core module) and forecasted neural signals as time elapses in the trial.

The NOSTIM agent, trained with a forecasting module but no stimulus vector, had performance that was almost entirely dependent on the core module (Fig. 5D). Removing the ability to forecast had almost no effect on performance (Fig. 5D, red line vs. green line). However, the NOSTIM agent’s performance dropped to chance levels when the core module was removed. These findings suggest that forecasting OFC activity does not help predict behavior unless task information is available to the agent. However, when additional task information is available, predicting behavior earlier within the trial is possible.

\section*{Discussion}
The purpose of neuro-engineering systems is to aid users in accomplishing their goals. Here we explored the use of economic value information in a hypothetical neuro-engineering system designed to interface with the user at the level of underlying motivations rather than actions. We demonstrate, using computation approaches with offline data, that the construction of a value-based system using RL approaches is feasible. Within a task in which a monkey is provided two options to choose from, we constructed an agent that could be trained to extract value-related information from the activity of neural ensembles in OFC and to then predict the monkey’s choice above chance, even in trials where there was no objectively superior option. We also demonstrated that such a system can be trained with neural and performance data from only a small number of trials (mean 126 per session), and that the system can be augmented to make faster predictions with the addition of a neural data forecasting module with access to sensory-related signals (e.g. from a hypothetical sensory assistive device) as part of a multi-modal or multi-source BMI architecture.

Not only did we observed the ability to make faster predictions when information about sensory stimuli is provided (Fig. 5C), but our analysis also revealed how important this information is to forecast neural activity. Forecasting neural data without providing external sensory-related signals offered negligible contributions to choice predictions early in a trial (Fig. 5D). Given the optimization procedure used (i.e. RL), the ablation analyses (Fig. 5C, D) showed that the agents trained with stimulus information learned to predict choice using information from the forecasting module rather than the core module (i.e. the actual neural activity). These findings suggest that forecasting OFC activity offers little behaviorally-predictive information unless informed by additional information (e.g. sensory variables).

\subsection*{Limitations and open questions }
Although we argue for using value signals in BMI applications, there are several limitations within our study. One limitation was our decision to not preserve the original trial order for training and subsequent testing. Although there are no multi-trial dependencies (i.e. stimuli are selected randomly), maintaining the original order of the trials would be a more realistic BMI scenario. In the current behavioral paradigm, there were only 12 unique stimuli and 132 unique combinations of first- and second-viewed stimuli, meaning that training on all possible trial types requires a relatively small number of trials.  In future studies we will explore decoders designed to predict choices on trials containing a stimulus familiar to the monkey but unobserved by the decoder during training, or on trials with stimuli novel to both the decoder and the monkey.

Along with trial ordering, we recognize that preprocessing using a training set, and the use of a specific temporal window (e.g. from 2nd target viewing) of neural activity are limitations. Because we require the decoder to identify the choice predictive signals within the neural activity, we surmise that the decoder could learn to make predictions without us needing to specify the relevant temporal window. The preprocessing approach used could be approximated and updated over the course of a session, however it is not clear how much those updates would affect model convergence. In future work, we seek to explore how less strict preprocessing steps affect decoding performance.

Within this study, we explored the performance capabilities of a decoder that had access to task-specific information in the form of a noiseless stimulus signal (for motivation, see the section 'Leveraging multi-source signals' below). Despite providing a noiseless stimulus signal, the system’s predictive performance was well below $100\%$, even for trials with large value differences between stimuli. Thus, our results could be viewed as a performance upper bound for making early-in-trial behavioral predictions using neural forecasting. These performance observations are very different from those reported in motor-based BMIs, where performance is nearly perfect (e.g., left vs. right reach). The difference in decoding performance between these brain areas may indicate important differences in encoding mechanisms between motor areas and prefrontal areas. Considering the reported properties of neural populations in the OFC (e.g. low-noise correlations \cite{conen2015neuronal},\cite{mcginty2023behavioral}), our observations may suggest the need to record from larger groups of cells. Alternatively, our lack of high decoding performance may suggest that other brain areas (e.g. amygdala \cite{stoll2024preferences}, posterior parietal cortex \cite{musallam2004cognitive}) may be more ideal for developing more robust value-based BMI.

Another limitation present in our study is our inability to assess how changes in the behavioral context affect decoder performance. In everyday settings, the relationship between sensory stimuli and subjective value can be highly context-dependent. For example, an umbrella would be highly valuable in certain contexts (a rainy day in the city or a sunny day at the beach) but low in others. Therefore, an important feature for a value-based BMI is the ability to use not only information about relevant stimuli, but also the behavioral context in which those stimuli appear. Furthermore, in natural settings, item values can change over time within the same context (e.g., sensory-specific satiety (\cite{rolls1986sensory})). Within a laboratory setting, we can manipulate the underlying mapping between stimulus features and value. Thus, an important question for future investigations is whether an agent equipped with an Internal Model could learn the appropriate stimulus-value mappings in differing contexts or could learn to change these mappings when needed.  For example, the values of the task stimuli in Fig. 1, are kept fixed in a given session. In future studies it would be possible to re-assign some stimulus to different values (or to no value at all), and to ask how quickly the agent’s choices adapt to reflect the updated values.

 \subsection*{An argument for the RL approach }
 The current behavior we consider is a simple form of goal-directed behavior which can be accomplished in a single stage by assigning values to two stimuli. However, in natural settings, goals must be pursued in more complex hierarchical situations that can have multiple steps and that can require making inferences using partial information (\cite{morris2018goal}, \cite{ribas2011neural},\cite{ribas2019subgoal}). In addition, goals can change according to context or internal demands. Thus, a successful BMI for goal-directed behavior must be flexible. It is for these reasons, along with the adaptability that the framework affords us, that we used an RL decoding approach. An RL approach allows for decoders that can facilitate a hierarchical viewpoint of complex movements like grasping an item (\cite{paul2019learning}, \cite{jurgenson2020sub}). In an analogous manner, hierarchical organization can be applied to goal-directed decision-making. The RL approach allows decoders to make choice predictions in task scenarios with intermediate steps and/or to adapt to changes of context and internal state. We acknowledge that in the relatively constrained behavioral paradigm we considered here that several supervised learning approaches could be used. Despite this limitation, we think that an RL decoding approach is ideal for augmenting goal-directed behavior because it is capable of the flexible inferences that underlie complex goal-based decisions in real-world settings. Future work will focus on using tasks with a hierarchical organization like those discussed in (\cite{schuck2016human}, \cite{ribas2019subgoal}) and exploring the assistive capabilities of such a neuro-engineering system.

\subsection*{Leveraging multi-source signals }
 We explored the case of an agent that has access to a module (Internal Model) that forecasts neural dynamics. Because in the current task value information in OFC is derived from visual stimuli, we also considered the case of forecasting OFC activity given an external signal that identifies the relevant stimuli. The motivation for this Internal Model is a hypothetical case in which the user is also assisted by a sensory prosthetic that samples the external environment to refine the agent’s computations. Here, we implement this assistance by providing the Internal Model a fixed vector uniquely defined for each stimulus, which assumes that the system had immediate access to noiseless information pertaining to a stimulus when the item was viewed by the monkey. We recognize that this implementation choice is a limitation when assessing the use of realistic stimulus signals by our model. Our formulation of the model allows the possibility to obtain stimulus information from a brain region upstream from the value computation process rather than an external source. In the behavioral paradigm considered this could entail retrieving stimulus information from a visual region like inferior temporal cortex (IT). Stimulus information in IT is typically evident at shorter latencies than value information in OFC (\cite{kar2018linking}), meaning that these signals could be used to render faster decisions than would be possible when relying on OFC alone.

 \subsection*{Utility of value signals in BMI applications }
Here, we consider the practical utility of a BMI that only has access to OFC activity. Unlike the constrained scenario we present in this study, a BMI for decoding economic choices in real-world settings would have to perform several inferences, e.g. not just the users’ desired option but the potential actions available to obtain that option. Although there may be passive scenarios that could be fulfilled decoding strictly from the OFC (\cite{alimardani2020passive}), the results from the agent with forecasting module (Figs. 4 and 5) shows the potential utility of OFC-decoded signals being paired with other regions. We will briefly outline two potential avenues illustrating the extremes of the active/passive spectrum of neuro-engineering devices. Here active neuro-engineering devices refer to those that operate based on explicit commands derived from the neural activity of the user. In contrast, passive devices rely on detecting changes in the user’s cognitive state.

One avenue is the pairing of value signals extracted from OFC activity with signals from motor or pre-motor cortex, to augment motor plans. For example, decision reaction times are faster when selecting high-value options (\cite{balewski2022fast}). This finding may highlight the use of value signals to adjust the speed of a robotic limb, computer cursor, or other BMI-driven effector. To our knowledge, OFC value signals have not been used to augment devices used in active BCI scenarios. However, Mussallam et al. (\cite{musallam2004cognitive}) used expected value signals from the parietal reach region to decode the desired location in a reach task in monkeys. In their case, the value signals in the parietal reach region were specific to the physical locations of the reach targets and were therefore used in an active BCI context. Signals related to economic value can be detected in numerous brain regions (\cite{maisson2021choice}, \cite{stoll2024preferences}, \cite{platt1999neural}, \cite{strait2015signatures}, \cite{yasuda2012robust}, \cite{sasikumar2018first}), which points towards the need to explore how these value signals differ in the information they encode and how they may be used to augment motor-based BCIs.

Another avenue is the pairing of value signals with regions that reflect other decision-related or cognitive signals. One domain that could leverage value signals is human-computer or human-robot interaction.  (\cite{alimardani2020passive}). For example, \cite{ehrlich2023human} demonstrated the use of cognitive signals to improve interactions between humans and robots, and (\cite{toichoa2021emotion}) used non-invasive approaches to decode a participant’s emotional state as it pertains to a robot’s actions.  In line with this construction, value signals could be combined with affective signals or error-related signals to guide the actions of a robot in collaborative tasks (\cite{staffa2023eeg}, \cite{toichoa2021emotion}). These studies suggest that there may be many cases in which abstract value signals can be used in neuro-engineering contexts to improve the interactions between users and assistive systems.

\bibliographystyle{plainnat}
\bibliography{refs}

\section*{Appendix}
\textbf{Table 1}. Statistics for Fig. 2C capturing the significance between performances when using Synth vs NoSynth data during training, using a paired t-test. Degrees of freedom DF = 22
\newline
\begin{tabular}{ |p{4cm}|p{3cm}|p{3cm}|p{3cm}|  }
 \hline
 \multicolumn{3}{|c|}{Table 1} \\
 \hline
 \textbf{Percentage of Training Trials} & \textbf{T-Statistic} &\textbf{P-Value}\\
 \hline
 80   & -1.89    &0.0707\\
70 & -0.68 & 0.5033 \\
60 & -0.64 & 0.5278\\
50 & 1.0 & 0.3282 \\
40 & -1.0 & 0.3282 \\
30 & -0.99 & 0.3282 \\
20 & 1.0 & 0.3282 \\
 \hline
\end{tabular}
\newline
\newline
\newline
\textbf{Table 2}. Statistics for Fig. 2D capturing the significance between performances when using synthetic data (Synth) vs not using synthetic data (NoSynth) during training, using a paired t-test. Degrees of freedom DF = 22
\newline
\begin{tabular}{ |p{4cm}|p{3cm}|p{3cm}|p{3cm}|  }
 \hline
 \multicolumn{3}{|c|}{Table 2} \\
 \hline
 \textbf{Percentage of Training Trials} & \textbf{T-Statistic} &\textbf{P-Value}\\
 \hline
80 &0.3 &0.7670 \\
70 &1.34 &0.1919 \\
60& 1.37 &0.1839 \\
50 &2.89 &0.0083 \\
40 &1.92 &0.0673 \\
30 &4.28 &0.0003 \\
20 &4.50 &0.0002 \\
\hline
\end{tabular}
\newline
\newline
\\
\textbf{Table 3}. Statistical significance for Fig. 3A (coefficient of determination relative to 0), using a paired t-test. Degrees of freedom DF = 22
\newline
\begin{tabular}{ |p{4cm}|p{3cm}|p{3cm}|p{3cm}|  }
 \hline
 \multicolumn{3}{|c|}{Table 3} \\
 \hline
 \textbf{Decoded Variable} & \textbf{T-Statistic} &\textbf{P-Value}\\
 \hline
Value of the 1st Target &3.25 &0.0036 \\
Value of the 2nd Target &2.28 &0.0326 \\
Value Difference between the 1st and 2nd Target &3.16 &0.0044 \\
\hline
\end{tabular}
\newline
\newline
\\
\textbf{Table 4}. Statistics for Fig. 4C capturing the statistical significance of the performance difference between the STIM vs NOSTIM case for the training data, using a paired t-test. Degrees of freedom DF = 22
\newline
\begin{tabular}{ |p{4cm}|p{3cm}|p{3cm}|p{3cm}|  }
 \hline
 \multicolumn{3}{|c|}{Table 4} \\
 \hline
 \textbf{Percentage of Training Trials} & \textbf{T-Statistic} &\textbf{P-Value}\\
 \hline
80 &3.15 &0.0046 \\
50 &3.13 &0.0048 \\
20 &3.15 &0.0046 \\
\hline
\end{tabular}
\newline
\newline
\\
\textbf{Table 5}. Statistics for Fig. 4D capture the statistical significance of the performance difference between STIM vs. NOSTIM cases for the held-out data, using a paired t-test. Degrees of freedom DF = 22
\newline
\begin{tabular}{ |p{4cm}|p{3cm}|p{3cm}|p{3cm}|  }
 \hline
 \multicolumn{3}{|c|}{Table 5} \\
 \hline
 \textbf{Percentage of Training Trials} & \textbf{T-Statistic} &\textbf{P-Value}\\
 \hline
80 &6.23 &2.79e-6 \\
50 &6.28 &2.54e-6 \\
20 &3.59 &0.0016 \\
\hline
\end{tabular}
\newline
\newline
\\
\textbf{Table 6}. Statistics for Fig. 5A. Statistical significance of the performance for each pair of models, using a paired t-test. Degrees of freedom DF = 22
\newline
\begin{tabular}{ |p{4cm}|p{3cm}|p{3cm}|p{3cm}|  }
 \hline
 \multicolumn{4}{|c|}{Table 6} \\
 \hline
 \textbf{Percentage of Training Trials} &\textbf{Comparison} & \textbf{T-Statistic} &\textbf{P-Value}\\
 \hline
80 &NOSTIM vs BASE &-3.79 &0.001 \\
 &NOSTIM vs STIM &6.03 &$4.48e-06$ \\
 &BASE vs STIM &2.40 &0.0250 \\
50 &NOSTIM vs BASE &-1.87  &0.0738 \\
 &NOSTIM vs STIM &4.56 &0.0002 \\
 &BASE vs STIM &2.86 &0.0090 \\
20 &NOSTIM vs BASE & -1.85 &0.0773 \\
 &NOSTIM vs STIM &2.91 &0.0081 \\
 &BASE vs STIM &2.52 &0.0192 \\
\hline
\end{tabular}
\newline
\newline
\\
\textbf{Table 7}. Statistics for Fig. 5B. Statistical significance was assessed using a paired t-test. Degrees of freedom DF = 22
\newline
\begin{tabular}{ |p{4cm}|p{3cm}|p{3cm}|p{3cm}|  }
 \hline
 \multicolumn{4}{|c|}{Table 7} \\
 \hline
 \textbf{Percentage of Training Trials} &\textbf{Comparison} & \textbf{T-Statistic} &\textbf{P-Value}\\
 \hline
80 &NOSTIM vs BASE &-1.86 &0.0756 \\
 &NOSTIM vs STIM &6.56 &$1.33e-06$ \\
 &BASE vs STIM  &8.28 &$3.31e-08$ \\
50 &NOSTIM vs BASE & -1.10 &0.2807 \\
 &NOSTIM vs STIM &8.42 &$2.49e-8$ \\
 &BASE vs STIM &8.50 &$2.12e-8$ \\
20 &NOSTIM vs BASE &-0.93 &0.3576 \\
 &NOSTIM vs STIM &3.46 &0.0022 \\
 &BASE vs STIM &2.58 &0.0168 \\
\hline
\end{tabular}
\end{document}